\documentstyle[aps,prb,multicol,epsf]{revtex}
\begin{document}
\draft
\title{Structural phase transition at high temperatures\\
in solid molecular hydrogen and deuterium}
\author{T. Cui$^{1,2}$, Y. Takada$^2$, 
Q. Cui$^1$, Y. Ma$^1$, G. Zou$^1$}

\address{$^1$National Lab for Superhard Materials, Jilin University, 
Changchun 130023, P. R. China\\
$^2$Institute for Solid State Physics, University of Tokyo, Kashiwa, 
Chiba 277-8581, Japan}
\date{28 Dec 2000}
\maketitle
\begin{abstract}
We study the effect of temperature up to 1000K on the structure 
of dense molecular {\it para}-hydrogen ({\it p}-H$_{2}$) and 
{\it ortho}-deuterium ({\it o}-D$_{2}$), 
using the path-integral Monte Carlo method. 
We find a structural phase transition from orientationally 
disordered hexagonal close packed (hcp) to an orthorhombic 
structure of {\it Cmca} symmetry before melting. 
The transition is basically induced by thermal fluctuations, 
but quantum fluctuations of protons (deuterons) are important 
in determining the transition temperature through effectively 
hardening the intermolecular interaction. 
We estimate the phase line between hcp and {\it Cmca} phases 
as well as the melting line of the {\it Cmca} solid. 
(accepted for publication in Physical Review B)
\end{abstract}
\begin{multicols}{2}
\narrowtext
\section{INTRODUCTION}
\label{sec:introduction}

The possibility for hydrogen to undergo a pressure-induced phase 
transition from a proton-paired insulator to a monatomic metal 
was first suggested in 1935. \cite{wigner35} 
Since then a large body of experimental and theoretical works 
have been done to determine its phase diagram in the pressure ($P$) 
and temperature ($T$) plane. 
Despite studies at $P$ reported to be as high as 342 GPa, 
\cite{narayana98} the tenacious covalent-bond feature of a pair of 
protons does not allow observation of a monatomic phase so far. 

For $P$ up to {\it ca.} 200GPa, the phase diagram has been 
investigated by several research groups 
at the room temperature and below, based on the optical 
measurement performed in the diamond anvil cell (DAC) devices. 
By now it is well established that the solid hydrogen exhibits 
at least three different molecular phases (the phases I$-$III), 
although some details are still in dispute. 
\cite{mao94,cui95,goncharov95,hemley96,mazin97,chen96,cui95b}
(1) Phase I: At $P < 110$GPa, each center of the H$_2$ molecule 
occupies the lattice site of an hcp structure, 
but quantum rotational effects overcome librational barriers, 
leading to an orientationally disordered phase. 
(2) Phase II or broken-symmetry phase (BSP): At $P$ between 110GPa 
and 150GPa, anisotropic intermolecular interactions freeze 
the molecular rotations into an orientationally ordered phase. 
(3) Phase III or H-A phase: At $P$ above 150 GPa, a third phase is 
observed, expected to be another kind of orientationally ordered 
phase. 

In 1996, Weir and coworkers gave one of the exciting results. 
They observed high electrical conductivity in the shock compressed 
H$_{2}$ and D$_{2}$ liquids and interpreted it as a transition 
from a semiconducting to a metallic diatomic fluid at 140 GPa 
and 3000K. \cite{weir96}
This experiment clearly demonstrates the importance of 
temperature effects in the search of metallization. 
For $T$ larger than 5000K, the effects are studied theoretically 
in the molecular, the dissociated, and the plasma regime of 
dense H$_{2}$. \cite{magro96,militzer00}
It is, however, still largely unclear how $T$ influences 
the state of solid H$_{2}$ at $T$ higher than the room temperature. 
We need to know the phase diagram at this temperature region 
to connect the liquid state with the low-temperature phases 
for a comprehensive understanding of dense H$_{2}$. 
Thus we focus our attention on the range of $P$ up to 200GPa 
and 300K$<T<1000$K in this paper. 

In order to study the condensed H$_2$ phases theoretically, 
several methods have been adopted at various levels of 
approximations to the {\em ab initio} Hamiltonian representing 
the coupled system of $N_a$ protons and $N_a$ electrons. 
These include the calculation of electronic energies 
in the local density approximation (LDA) or its refinements 
to the density functional theory for a variety of 
crystal structures with molecular orientations fixed to 
some particular configuration, 
\cite{a68,f77,b89,k91,k92,n92,s93,e96,s97,n98,n99,j00,s00} 
and the implementation of {\em ab initio} molecular dynamics 
treating protons 
as classical particles.\cite{kohanoff95,kohanoff97,kohanoff99} 
We should note, however, that the strong quantum nature of 
light protons requires a more careful quantum-mechanical 
description of their zero-point motion. 
In fact, using either the first-principle path-integral
molecular dynamics \cite{kitamura00} 
or the quantum Monte Carlo (QMC) simulations, 
\cite{magro96,ceperley87,natoli95} 
the calculation treating protons as dynamic quantum particles 
has already been performed. 
Although desirable, this approach demands the computer resources 
very much. 
Thus, even in a recent paper, \cite{kitamura00} 
the calculation is done only with as small as $N_a=64$. 
This size of $N_a$ is too small in order to obtain a correct 
equilibrium structure free from any bias of an initially 
assumed one through simulations. 

A ten times increase in $N_a$ is possible if we approach the 
problem by adopting the hydrogen molecule as a basic ingredient 
in simulations rather than the proton-electron mixture. 
In this approach, the system is reduced to a quantum-mechanical 
problem of $N$ (=2$N_a$) molecules interacting to each other 
through an effective intermolecular pair potential $V_{\rm pair}$. 
Thus the QMC calculation is implemented only for the nuclear 
degrees of freedom 
\cite{zoppi91,runge92,wagner94,kaxiras94,zoppi95,cui97} 
and the electronic degrees of freedom are implicitly taken 
into account in the choice of $V_{\rm pair}$. 
In the present paper, we adopt this approach and employ 
the finite-temperature path-integral Monte Carlo (PIMC) method 
to investigate both lattice and orientational transitions in the 
molecular phase with zero-point motions incorporated rigorously. 
We shall use an empirically determined $V_{\rm pair}$ 
which is the sum of the Hemley-corrected Silvera-Goldman 
potential \cite {hemley90} and the Runge-scaled Shaefer 
potential. \cite{runge92} 
This choice of $V_{\rm pair}$ is known to give an equation 
of state in very good agreement with experiment 
as well as the I/II phase boundaries for both H$_2$ 
and D$_2$ over the whole experimentally investigated range of 
pressures. \cite{cui97}

This paper is organized as follows: 
We shall explain our theoretical model in more detail 
in Sec.\ \ref{sec:model}, including a review of trials 
to determine $V_{\rm pair}$, a description of our choice of 
$V_{\rm pair}$, a brief summary of the PIMC method with a 
constant-pressure ensemble, and some computational details. 
In Sec.\ \ref{sec:results} our results are shown for 
both solid H$_{2}$ and D$_{2}$, indicating a temperature-induced 
solid-solid structural phase transition. 
The structure of the new high-temperature phase, the solid-liquid 
phase transition, and the phase diagram are also given here. 
Section \ref{sec:discussion} contains a discussion on the role 
of zero-point motion of protons in the solid-solid structural 
phase transition and finally in Sec.\ \ref{sec:summary} we 
summarize our results. 

\section{THEORETICAL MODEL}
\label{sec:model}
\subsection{Molecule base}

Within the temperature and pressure range that a molecule can 
be regarded as a basic ingredient, a quantum molecular solid 
can be described by the Hamiltonian $H_{\rm nuclear}$ as 
the sum of $T_{\rm nuclear}$ and $V_{\rm nuclear}$, where 
\begin{eqnarray} 
\label{eq:hamiltonian}
T_{\rm nuclear}= -\frac{\hbar^2}{2m}\sum_i^N\nabla_{R_i}^2 
+\frac{\hbar^2}{2I}\sum_i^N{\bf L}_i^2, 
\end{eqnarray} 
and
\begin{eqnarray} 
V_{\rm nuclear}&=&\frac{1}{2}\sum_{i\neq j}^N
V_{\rm pair}({\bf R}_{ij},{\bf\Omega}_i, {\bf\Omega}_j). 
\end{eqnarray} 
Here ${\bf R}_i$ is the center-of-mass position vector 
of the $i$th molecule, ${\bf \Omega}_i$ its orientation vector, 
${\bf L}_i$ its angular momentum operator, 
and the intermolecular separation vector 
${\bf R}_{ij}\equiv {\bf R}_i-{\bf R}_j$. 
The molecular mass and moment of inertia are denoted 
by $m$ and $I$, respectively. For a linear molecule with the 
intramolecular bond length fixed, $I$ is a scalar constant.
The values of $m=3676$ and $7352$ (atomic units) for H$_2$ 
and D$_2$, respectively, and rotational constants 
$B\equiv\hbar^2/2I= 84.98$ 
and 42.92 K for H$_2$ and D$_2$, respectively, 
are used. \cite{runge92,silvera80}
Among the the nuclear degrees of freedom, we include both 
translation and rotation modes of molecules 
in the kinetic energy $T_{\rm nuclear}$, but we consider 
implicitly the intramolecular vibration mode along with the 
electronic degrees of freedom by judiciously choosing 
$V_{\rm pair}$ the pair-wise sum 
of which gives the potential energy $V_{\rm nuclear}$. 

We have enough evidence to support the present approach 
in our research range ($0\sim200$GPa and $0\sim1000$K). 
First, we can neglect the molecular dissociation, 
because its fraction is less than 0.01\% from the same 
analysis as the one leading to about 5\% at 140GPa 
and 3000K. \cite{weir96,holmes95}
Secondly, {\em ab initio} quantum-chemistry calculations 
at $N$=2 lead us to the conclusion that the dependence 
of $V_{\rm pair}$ on the bond length $r_0$ is weak, 
even though $r_0$ shrinks by more than 10\% from 1.4 bohr (0.74\AA) 
as two H$_2$ molecules come closer. \cite{ree79} 
This assures the validity to employ an $r_0$-independent 
$V_{\rm pair}$. The notion of $V_{\rm pair}$ is relevant 
if $r$ the intermolecular distance is 
larger than about 2.3 bohr (1.2\AA), 
because each molecule is estimated to possess a repulsive core 
of the radius of 0.6\AA. \cite{magro96}
Finally the effect of zero-point intramolecular vibrations 
(vibrons) can be taken into account in the form of $V_{\rm pair}$ 
as we shall illustrate in Sec.\ \ref{sec:discussion}
by using a toy model. 

\subsection{H$_2$-H$_2$ interaction}
\label{sec:potential}

\begin{figure}[h]
\centerline{\epsfxsize=2.5truein \epsfbox{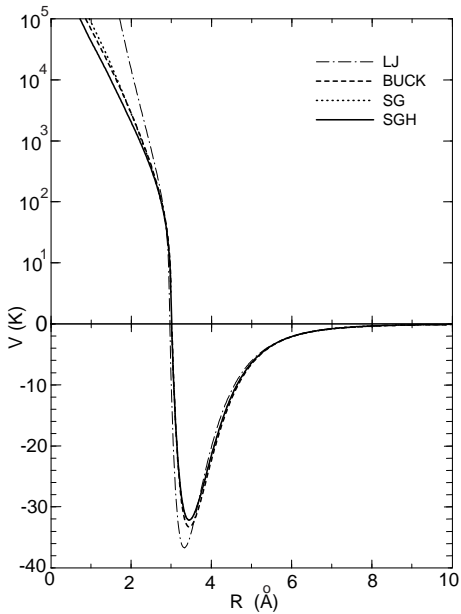} }
\caption{Various potentials proposed for the interaction 
between two H$_2$ molecules. 
The solid, dotted, dashed, and dotted-dashed curves represent, 
respectively, the SGH, SG, BUCK, and LJ potentials.} 
\label{fig:isopot}
\end{figure}

Rather than determining $V_{\rm pair}$ from first principles, 
we shall resort to a phenomenological approach which is well 
reviewed in the literature. \cite{silvera80,buck83} 
In general, the multi-dimensional function 
$V_{\rm pair}({\bf R}_{12},{\bf \Omega}_1,{\bf \Omega}_2)$ 
can be expressed as the sum of the isotropic pair 
potential $V_{0}(R_{12})$ and the anisotropic potential 
$V_{ani}({\bf R}_{12},{\bf \Omega}_1,{\bf \Omega}_2)$. 
A variety of analytic forms have been proposed 
for $V_0(R)$ and their relevance has been tested 
in the past. Four of them are illustrated 
in Fig.~\ref{fig:isopot}. 
Because the Lennard-Jones (LJ) two-parameter form (the 
dotted-dashed curve in Fig.~\ref{fig:isopot}) has an 
unphysically strong repulsive core, more sophisticated forms 
have been proposed. 
The most commonly used form is either 
BUCK potential $V_{\rm BUCK}$ \cite{buck83} or SG 
potential $V_{\rm SG}$, \cite{silvera78} plotted 
by dashed and dotted curves, respectively, 
in Fig.~\ref{fig:isopot}. 
Although they have almost the same analytic form, 
$V_{\rm BUCK}$ is fitted to the gas-phase data, 
while $V_{\rm SG}$ to the solid-phase data. 
In order to include the three-body interactions effectively 
in the solid environment, a further refinement is introduced 
to the original form of $V_{\rm SG}$ by incorporating an additional 
repulsive term, leading to the final form of SG potential. 
Compared to $V_{\rm BUCK}$, the net effects contributed from
the three-body interaction are a hardening of $V_{\rm SG}$
at small $R$ and a slight raising of the well depth. 
Their concrete forms can be found in 
Refs.~\onlinecite{cui97,silvera80}, and \onlinecite{buck83}. 

This $V_{\rm SG}(R)$ works well for the solid H$_2$ and D$_2$ 
under ambient pressures. 
Under high pressures, however, an additional softening effect 
found by Hemley {\em et al}. is needed to correctly describe 
the enhanced many-body contributions in the short-range region 
caused by the dense solid environment. 
This effect can be incorporated by an {\em ad hoc} 
short-range correction $V_{\rm SR}(R)$ to 
$V_{\rm SG}(R)$, \cite{hemley90} 
leading to the definition of the Hemley-corrected SG potential 
$V_{\rm SGH}(R)$ (the solid curve in Fig.~\ref{fig:isopot}), 
where the actual form of $V_{\rm SR}(R)$ can be found in 
Refs.~\onlinecite{cui97} and \onlinecite{hemley90}. 
This softening correction at small $R$ is very important 
in reproducing the equation of state (EOS) 
derived experimentally from the liquid D$_2$ 
shock-wave data, \cite{nellis83,ross83} 
as assessed by Cui {\em et al}. in Ref.~\onlinecite{cui97} 
in which the EOS for D$_2$ at $T=20$K and for $P$ up to 40GPa 
is calculated for testing the intermolecular potential. 
From the test, the EOS is proved to be insensitive 
to the details of $V_{ani}$, 
but it is very sensitive to the choice of $V_{0}(R)$ 
which controls the translational motion of molecules. 
In fact, $V_{\rm SGH}(R)$ gives results in excellent agreement 
with the experimental data, \cite{hemley90} 
while $V_{\rm SG}(R)$ does not. 
The same conclusion is also drawn at $T=300$K 
from a similar test; 
for $P$ up to 200GPa, $V_{\rm SGH}(R)$ provides the results 
for the EOS in line with the experimentally obtained 
fitting formula. \cite{cui97}

Only a few works have been done for the assessment of $V_{ani}$. 
Using the {\em ab initio} quantum chemical method to evaluate 
the contributions such as the long-range electronic 
quadrupole-quadrupole interaction and the atom-diatom
scattering term, Schaefer {\em et al}. have given a potential 
for $V_{ani}$. \cite{schaefer79,monchick80,schaefer89} 
Comparing the results obtained from the electronic-structure 
calculation in the LDA, Runge {\em et al}. have found that 
this Schaefer potential $V_{\rm Schaefer}$ is too repulsive 
in the dense solid. 
Thus, in order to soften it, they have introduced a scaling factor 
to $V_{\rm Schaefer}$ to provide $V_{ani}$ as 
$\alpha_n V_{\rm Schaefer}$. \cite{runge92} 
This factor $\alpha_n$ is linear in the nearest neighbor spacing 
$R_{nn}$ to correctly describe the dense-solid environment as 
$\alpha_n\!=\!0.61\!+\!0.31(R_{nn}/R_{nn}^0\!-\!0.5)$ with 
$R_{nn}^0\!=\!3.789$\AA, the H$_2$ equilibrium zero-pressure 
nearest-neighbor spacing. 
Both the bare $V_{\rm Schaefer}$ 
and the scaled one for $V_{ani}$ have 
been tested in Ref.\ \onlinecite{cui97} 
to find that this difference is crucial in reproducing 
the I/II phase line for the solid D$_2$. 
The best agreement with the experimental data for $P$ up 
to 150GPa is achieved by using the scaled 
anisotropic potential combined with $V_{\rm SGH}(R)$ for $V_0(R)$. 
The bare $V_{\rm Schaefer}$ predicts the transition to occure 
at much lower pressures. 
This difference was also observed in the fixed-lattice 
PIMC study. \cite{runge92} 

For those reasons mentioned above, we shall choose the sum of 
$V_{\rm SGH}(R)$ and the scaled Schaefer potential 
as $V_{\rm pair}$ in this paper. 
We believe that this $V_{\rm pair}$ is most reliable 
for the study of solid molecular H$_2$ and D$_2$ 
in our research range, in particular, for $P < 150$GPa. 

\subsection{Path Integral Monte Carlo method with 
constant pressure ensemble}
\label{sec:pimc}

With the Hamiltonian thus provided, we model the bulk solid by a 
simulation cell, which is periodically duplicated in all three 
spatial dimensions to minimize surface and finite-size effects. 
Initial size and geometry of the simulation cell are chosen 
to accommodate a particular number of molecules ($N$) 
and hcp structure. 
Our calculations are performed mostly on $N$ =288 
and from an initial simulation cell determined by 
two basis vectors ({\bf a}$_p$ and {\bf b}$_p$) 
forming a 60$^0$ angle and by the third one ({\bf c}$_p$) 
perpendicular to them with the length ratio of 
$a_p:b_p:c_p=1:1:4\sqrt{6}/9$. 
The packing pattern is ABAB... to form the hcp lattice structure. 
Extensive testing on $N$ = 96, 150 and 392 has also been made.

In studying the effect of heating solid H$_2$ isobarically 
at high pressures, we perform the path-integral Monte Carlo 
calculations with a constant-pressure ($NPT$) ensemble, 
instead of a simpler constant-volume ($NVT$) one, 
to avoid a bias of the restrictive cell geometry 
with a predetermined crystal structure. 
Implementation of the $NPT$ ensemble is achieved 
by extra independent Metropolis moves of 
three basis vectors {\bf a}$_p$, {\bf b}$_p$ and {\bf c}$_p$, 
which generates a Markov chain of states having a limiting 
distribution proportional to $\exp(-H/T+N\ln V)$ 
at $T$ with the enthalpy $H$($=PV+E_s$), where $E_s$ is 
the energy expectation value $\langle H_{\rm nuclear} \rangle$ 
of the configuration $s$ with $s$ representing a set of 
scaled coordinates. 
This enables us to monitor volume changes and therefore 
to observe a possible first-order phase transition directly. 

In order to avoid the ``minus-sign problem'' in QMC studies 
inherent in fermions, we confine ourselves 
to studying {\it p}-H$_2$ and {\it o}-D$_2$. 
After 4000 Monte Carlo steps for equilibration, 
statistical averages are collected from every second step, 
to a total of about 4000 data points. 
Because we focus on higher temperatures than the room temperature, 
relatively smaller partition number in the imaginary-time axis $(M)$ 
is enough. 
At a fixed $P$, we start with simulations at the room temperature 
at which $M$ is increased up to 16 to ensure the convergence. 
We come to know that $M =8(5)$ is enough for $P=30$GPa (higher $P$). 
Once its large enough value is determined, 
$M$ is held fixed during heating at the given $P$. 
This allows us to use the equilibrium configuration 
at low temperature as a start-up one for higher temperature, 
resulting in continuous heating of our sample. 
Because other computation techniques are documented in 
Ref.\ \onlinecite{cui97}, we omit further details of them here.

\section{RESULTS}
\label{sec:results}

We observe the equilibrium structure in real space directly 
and also monitor the pair distribution function $g(R)$ 
that represents the conditional probability of 
finding other molecules at the distance $R$ 
from a specified molecule at the origin:
\begin{equation}
g(R)=\frac{1}{4\pi R^2\rho}\left\langle
\sum_{i<j}\delta(R_{ij}-R) \right\rangle, 
\label{eq:gofr} 
\end{equation}
with $\rho$ the density of system. 
This function shows a set of well-defined peaks characteristic 
to the configuration of neighboring molecules around the one 
at the origin. 
The function $O(R)$ defined in Ref.\ \onlinecite{cui97} 
is used to measure the correlation in molecular orientation. 

\begin{figure}[h]
\centerline{\epsfxsize=3.4truein \epsfbox{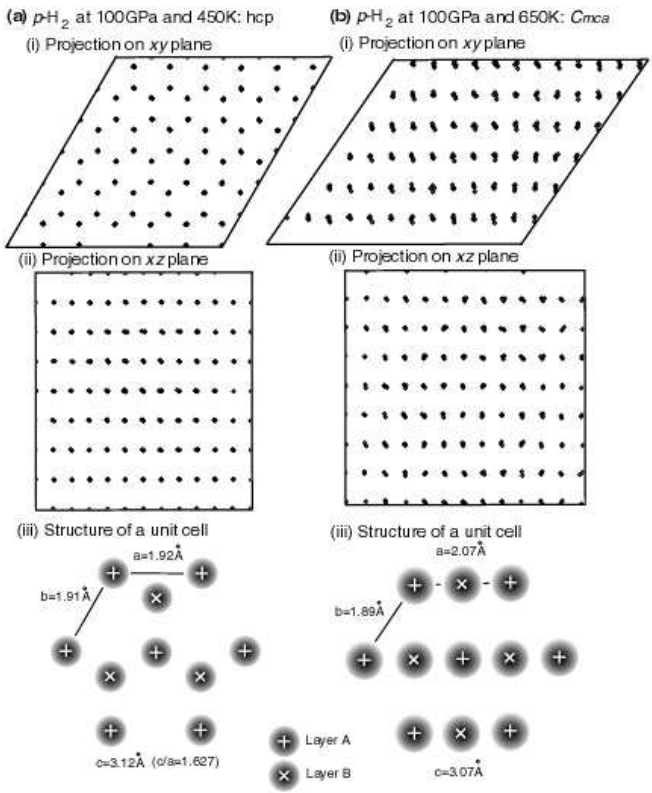} }
\caption{Snapshots of the average equilibrium distribution 
of molecules at 100GPa in real space projected on 
both (i) $xy$ and  (ii) $xz$ planes. 
Row (a): {\it p}-H$_2$ at 450K; row (b): {\it p}-H$_2$ at 650K. 
The structure of each unit cell is shown in (iii) 
in which the size of fluctuations is represented 
by the painted region.} 
\label{fig:snapshot}
\end{figure}

The equilibrium distribution of 288 ($=6 \times 6 \times 8$) 
molecules in real space projected on both $xy$ and $xz$ planes 
are shown in Fig.~\ref{fig:snapshot} 
for {\it p}-H$_2$ at 100GPa and at two different temperatures. 
Although it fluctuates with the standard deviation 
indicated by the painted region in Fig.~\ref{fig:snapshot}(iii), 
on the average, each molecule occupies the lattice site of hcp 
at 450K, while that of an orthorhombic {\it Cmca} structure at 650K. 

\begin{figure}[h]
\centerline{\epsfxsize=2.9truein \epsfbox{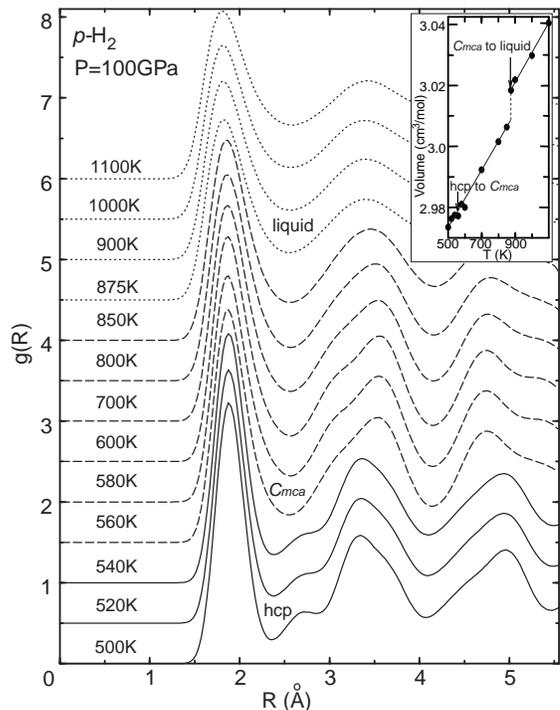} }
\caption{Temperature dependence of $g(R)$ for {\it p}-H$_2$ 
at 100GPa. 
Inset shows the temperature dependence of volume. 
(Statistical errors are smaller than the size of symbols.)}
\label{fig:gr}
\end{figure}

For detecting this transition quantitatively, 
we plot $g(R)$ at $P=100$GPa with the increase of $T$ 
in Fig.~\ref{fig:gr}. 
Below 560K the second peak in $g(R)$ at $R\sim2.72$\AA, 
a characteristic feature of hcp, is clearly seen. 
But the peak disappears at higher $T$, 
which is a sign of the structural transition from hcp to {\it Cmca}. 
Thus we identify the transition temperature $T_{tr}$ as 560K 
with a statistical error of 20K. 
(By ``statistical'' we mean that the difference of states cannot 
be seen clearly if that in $T$ is less than 20K 
due to statistical fluctuations.) 
This structural change is always seen in the system with $N=$ 96, 
150, or 392. 
We find a large $N$ dependence of $T_{tr}$, 
leading to an error of 40K or larger incurred 
at extrapolation of $T_{tr}$ at $N \to \infty$. 
This error cannot be reduced further, 
because calculations with a much larger $N$ 
are not feasible at present. 

With the further increase of $T$ from $T_{tr}$, 
the peaks in $g(R)$ become less sharp. 
Eventually at $T$ above 875K (which we identify 
the melting temperature $T_m$ with a statistical error of 25K, 
about the same size of an error due to extrapolation 
in $N \to \infty$) 
$g(R)$ exhibits characteristics of a liquid phase. 
This first-order melting transition can be much better identified 
by the discontinuous change with $T$ in either the enthalpy $H$ 
or volume $V$ (the inset of Fig.~\ref{fig:gr}).

\begin{figure}[h]
\centerline{\epsfxsize=3.55truein \epsfbox{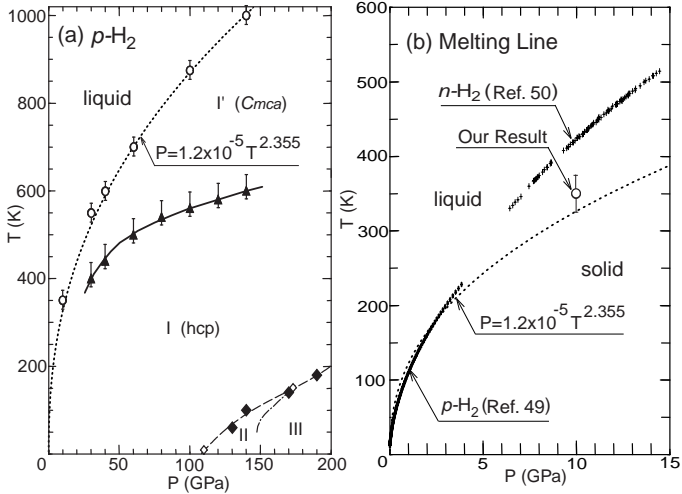} }
\caption{(a)Phase diagram of {\it p}-H$_2$. 
At low temperatures, the dashed curve shows the I/II phase boundary 
interpolated through the two measured data marked 
by open diamonds (Refs.\ \protect\onlinecite{cui95} 
and \protect\onlinecite{mazin97}), 
while the dotted-dashed curve indicates 
the experimental II/III phase boundary. 
Solid diamonds represent the orientational order-disorder 
phase transition points obtained by PIMC in 
Ref.\ \protect\onlinecite{cui97}. 
(Size of the symbols represents the magnitude of errors.) 
Solid triangles and open circles for the hcp/{\it Cmca} transition 
and the melting, respectively, show the new phase boundaries 
obtained in this study. 
The dotted curve is a Simon fit to the melting line of this work. 
(b) Melting line at low pressures with experimental 
data for {\it n}-H$_2$ (pluses)\protect \cite{datchi00} 
and for {\it p}-H$_2$ (stars). \protect \cite{driessen79}}
\label{fig:phased}
\end{figure}

\begin{figure}[h]
\centerline{\epsfxsize=3.4truein \epsfbox{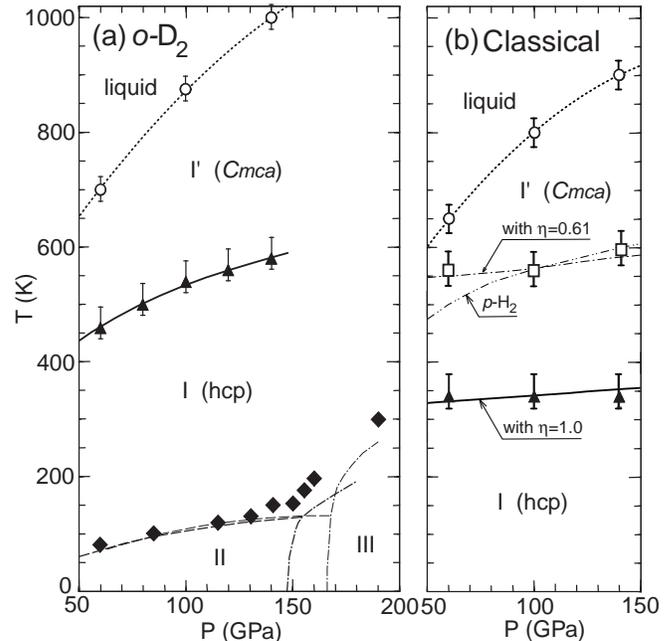} }
\caption{(a) Similar phase diagram for {\it o}-D$_2$ with symbols 
indicating just the same meanings as those 
in Fig.~\ref{fig:phased}(a). 
Two sets of experiments have detected three distinct 
phases I, II, and III for $P$ below 200GPa at low temperatures, 
as indicated by two dashed and dotted-dashed curves 
with different thickness, 
taken from Refs.\ \protect\onlinecite{mao94} 
and \protect\onlinecite{mazin97}. 
In (b), solid triangles are the corresponding high-temperature 
phase transition points for the classical reference system ($M=1$),
while open squares show the transition points by the pair potential 
with $\eta = 0.61$, defined in Eq.\ (\ref{eq:ourv1}). 
(See the text in Sec.\  \ref{sec:discussion}.) 
For comparison's sake, the double-dotted-dashed curve shows the 
data for solid H$_2$, the same as in Fig.~\ref{fig:phased} (a).}
\label{fig:referphased}
\end{figure}

By evaluating both $T_{tr}$ and $T_m$ at other pressures, 
we obtain the phase diagram of dense {\it p}-H$_2$ 
in the $T-P$ plane as shown in Fig.~\ref{fig:phased}(a). 
The new solid phase is labeled as phase I'. 
Since $O(R)$ vanishes, its structure is orientationally disordered. 
The I'/I and liquid/I' phase boundaries are indicated 
by solid triangles and open circles with error bars, respectively. 
Large curvature of the phase line for $P < 50$GPa reflects 
the rapid change of the volume in that region. 
A Simon equation of $P=1.2 \times 10^{-5} T_m^{2.355}$ 
(the dotted curves in Fig.~\ref{fig:phased}) 
fits our results rather well. 
A lack of experiment on $T_m$ at high pressures makes 
a precise comparison between theory and experiment difficult, 
but we obtain a reasonably good agreement with the 
data \cite{driessen79} for $P < 4$GPa, as can be seen in 
Fig.~\ref{fig:phased}(b). 
The data for {\it n}-H$_2$ up to 15GPa, \cite{datchi00} are 
always higher than our results, reflecting the fact that 
the presence of {\it o}-H$_2$ makes $T_m$ higher. \cite{driessen79}
In Fig.~\ref{fig:referphased}(a) we show our result 
for dense {\it o}-D$_2$ exhibiting a similar phase diagram. 
The difference in $T_{tr}$ between these two systems 
becomes smaller with the increase of $P$, but {\it p}-H$_2$ gives 
definitely a higher $T_{tr}$ than {\it o}-D$_2$. 
As for $T_m$, we find no meaningful difference. 
\section{QUANTUM HARDENING EFFECT}
\label{sec:discussion}

For understanding the mechanism to bring about the phase I', 
we have performed another simulation with $M=1$ 
(the ``classical reference'' system in which the molecules are 
treated as classical particles) and found an analogous phase 
diagram as shown in Fig.~\ref{fig:referphased}(b). 
In this system, only the temperature-induced random motion 
of molecules can play a role in transforming a compact hcp structure 
into a less compact {\it Cmca} one and finally into a liquid. 
Thus we can identify the thermal fluctuations 
as the basic driving force of the present phase transition. 

\begin{figure}[h]
\centerline{\epsfxsize=3.4truein \epsfbox{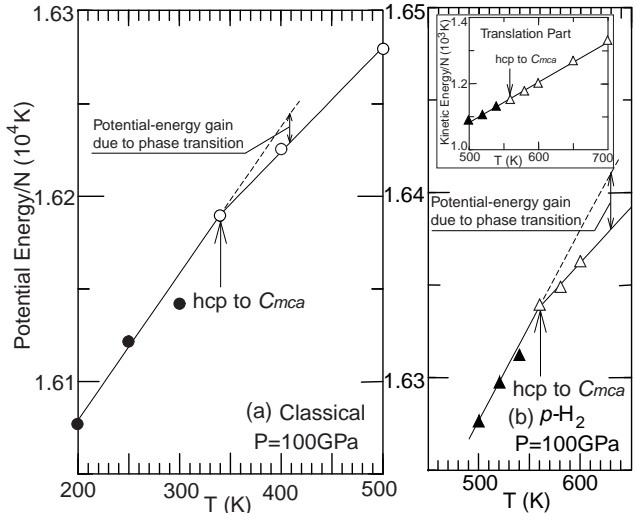} }
\caption{Temperature dependence of the potential energy 
$\langle V_{\rm nuclear} \rangle$ per molecule for both (a) the 
classical reference system and (b) {\it p}-H$_2$ 
near respective $T_{tr}$'s at 100GPa. 
Data in hcp and {\it Cmca} structures are shown by solid and open 
symbols, respectively. The inset in (b) plots the change in 
the translational kinetic energy per molecule.}
\label{fig:potential}
\end{figure}

Examining the changes of various physical quantities at $T$ 
around $T_{tr}$, we found that the most important change occurs 
in the potential energy. 
The energy increases with $T$ in both the classical 
(Fig.~\ref{fig:potential}(a)) and the H$_2$ 
(Fig.~\ref{fig:potential}(b)) systems due primarily to the 
fluctuation of molecules to shorten the nearest-neighbor distance 
$R_0\sim1.9$\AA, because $V_{\rm pair}$ is very repulsive 
at $R$ near $R_0$ and it changes very rapidly with $R$. 
(See Fig.~\ref{fig:isopot}.) 
We should note that the increasing rate is larger in the compact 
hcp phase than that in the less compact {\it Cmca} phase. 
In this regard, the phase transition is induced 
by the potential-energy gain as indicated in 
Fig.~\ref{fig:potential}. 
No such gain is seen for the translational kinetic energy 
as shown in the inset of Fig.~\ref{fig:potential}(b).

Eventually the reason why $T_{tr}$ in the classical system is much 
lower than that in solid {\it p}-H$_2$ must be ascribed to 
the quantum zero-point motion of protons, the only ingredient 
that is not included in the classical system. 
Such a motion is, in a sense, regarded as a kind of vacuum 
polarization the effect of which, in the context of high-energy 
physics, can be renormalized into fundamental physical quantities 
such as mass and interaction constant. 
We envisage that this must also be the case here. 
In order to substantiate this, we consider a toy model, 
namely, an interacting two-particle system in a harmonic potential 
in one dimension, described by the Hamiltonian as 
\begin{equation}
H={p_1^2 \over 2m}+{Kx_1^2 \over 2}
+{p_2^2 \over 2m}+{Kx_2^2 \over 2}+V(x_1-x_2), 
\label{eq:toyh} 
\end{equation}
from which we can derive the Hamiltonian for the relative 
motion as 
\begin{equation}
H_r={p^2 \over 2\mu}+\mu\omega^2x^2+V(x), 
\label{eq:toyrel} 
\end{equation}
where $\mu=m/2,\omega=\sqrt{K/2\mu}$ and $x=x_1-x_2$. 
By writing $\psi(x)=\psi_{0}(x)\varphi(x)$ and 
$E=\omega/2+\varepsilon$ 
with $\psi_{0}(x)=\sqrt[4]{\mu\omega/\pi}\exp(-\mu\omega x^{2}/2)$, 
we can cast Eq.~(\ref{eq:toyrel}) into 
\begin{equation}
\left ({p^2 \over 2\mu}+\widetilde{V}(x)
\right )\varphi(x)=\varepsilon\varphi(x), 
\label{eq:toyfinal}
\end{equation}
where the effective interaction $\widetilde{V}(x)$ is related 
to the bare one through 
\begin{eqnarray} 
\widetilde{V}(x) \equiv V(x)-\mu^{-1}{\psi_0^{'} \over \psi_0}
{\varphi^{'} \over \varphi}
=V(x)+\omega x {\varphi^{'} \over \varphi},
\label{eq:effectv} 
\end{eqnarray} 
We see that the effect of the zero-point oscillation is 
included effectively in $\tilde{V}(x)$, 
implying that the physics can be captured in terms of 
$\tilde{V}(x)$. \cite{anharmonicity} 
In this sense, $V_{\rm pair}$ can include the effect of 
intramolecular vibrons. 

In general the closed-packed structure is known to become more 
stable with a harder $V_{\rm pair}$. \cite{pettifor95}
In the above toy model, we readily know that, if $V(x)$ is 
repulsive, $x\varphi^{'}/\varphi$ is positive, 
leading $\widetilde{V}(x)$ to be more repulsive than $V(x)$. 
This indicates that the zero-point motion of protons can make 
$V_{\rm pair}$ effectively more repulsive than that in the 
classical system in which $\widetilde{V}(x)$ is reduced to 
$V(x)$ owing to $\omega=\sqrt{K/m} \to 0$ as $m \to \infty$. 
\begin{figure}[h]
\centerline{\epsfxsize=3.2truein \epsfbox{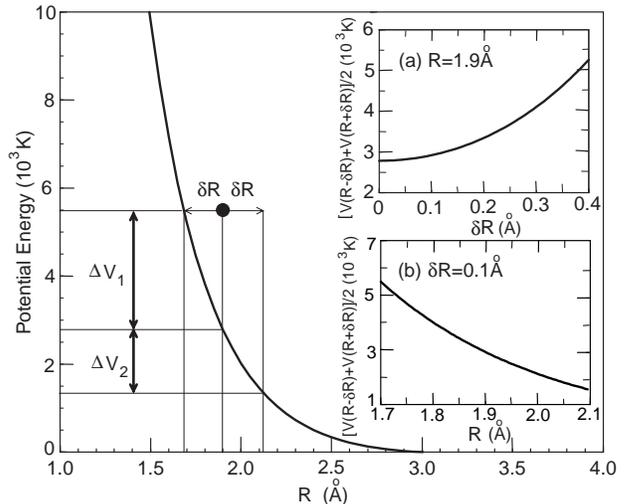} }
\caption{Schematic view to understand the Fluctuation effect 
on the potential energy. 
Inset: (a) average potential energy as a function of fluctuation 
$\delta R$ with $R = 1.9$ \AA. 
(b) average potential energy as a function of separation $R$ 
with $\delta R = 0.1$ \AA.}
\label{fig:fluc}
\end{figure}

This hardening effect due to the molecular fluctuations 
can be understood more visually through Fig.~\ref{fig:fluc}. 
At the pressure range considered in this paper, the repulsive 
part of the interaction plays a central role in determining the 
potential energy. 
The fluctuation shortens one distance between the neighboring 
molecules, but at the same time it lengthens another one. 
Thus we have to average these two opposite effects. 
Because the potential changes very rapidly with $R$ the 
intermolecular distance, $\Delta V_1$ the increase of the potential 
due to the shortening is larger than $\Delta V_2$ the decrease 
due to the lengthening. 
Therefore the average potential energy, 
$[V(R-\delta R)+V(R+\delta R)]/2$, becomes larger than $V(R)$ 
the potential energy without the fluctuation. 
As shown in the inset (a) of Fig.~\ref{fig:fluc}, the larger 
the fluctuation, the more repulsive the potential energy. 
This explains the behavior of the potential energy 
in Fig.~\ref{fig:potential}; 
the solid H$_2$ has a larger potential energy than the classical 
reference system at the same pressure and temperature, 
because it provides larger fluctuations due to the combination 
between the quantum and thermal effects. 
This also manifests itself in the much larger potential-energy gain 
in Fig.~\ref{fig:potential}(b) than that in 
Fig.~\ref{fig:potential}(a).
Thus we can conclude that the molecules in the quantum system 
feel an effectively more repulsive potential 
than those in the classical system. 
This ``quantum hardening effect'' of $V_{\rm pair}$ explains 
why hcp becomes more stable in {\it p}-H$_2$. 

With the increase of pressure, the deviation of the I/I' phase 
lines between solid H$_2$ and the classical system is getting 
larger. 
In order to understand this result, we plot 
$[V(R-\delta R)+V(R+\delta R)]/2$ as a function of $R$ 
with a given $\delta R$ in the inset (b) of Fig.~\ref{fig:fluc}. 
The average potential energy increases with the decrease 
of $R$, a behavior of $R$ seen with the increase of pressure. 
Thus we conclude that the quantum hardening effect becomes 
stronger with increasing pressure and this explains 
the above result well. 

In order to study this quantum hardening effect more 
quantitatively, we have performed another simulation 
in the classical system with the pair potential, defined as 
\begin{equation} 
V_{\rm pair}=V_{\rm SG} + \eta V_{SR} + V_{ani}, 
\label{eq:ourv1} 
\end{equation} 
where the parameter $\eta$ is introduced to measure the softness 
of $V_{\rm pair}$; $\eta = 0$ for the hardest SG potential
and $\eta = 1$ for the softest SGH one. 
In Fig.~\ref{fig:referphased}(b), the open squares with error 
bars show the I/I' phase boundary obtained using 
$V_{\rm pair}$ with $\eta = 0.61$. 
Notice that this more repulsive potential than the SGH one 
makes the I/I' phase boundary of the reference classical system 
almost the same as the one for the quantum solid H$_2$ with 
the SGH potential (the double-dotted-dashed curve) at $P=100$GPa. 
We need a softer (harder) $V_{\rm pair}$ to better fit the data at 
lower (higher) pressure, but this behavior is consistent with the 
above-mentioned pressure dependence of the quantum hardening 
effect. 

\section{SUMMARY}
\label{sec:summary}

Utilizing an intermolecular interaction potential $V_{\rm pair}$ 
proved to be reliable at high pressures in the PIMC simulations, 
we have found the structural phase transition from 
orientationally disordered hcp to $Cmca$ induced by thermal 
fluctuations in both dense solid H$_2$ and D$_2$ before melting. 
The result shows that a new phase exists at high temperatures 
in the solid molecular systems. 
The potential-energy gain due to the enhancement of fluctuations 
with increasing temperature is the main source to bring about 
the structural phase transition. 
The strong quantum zero-point motion plays an important role 
in determining $T_{tr}$ the transition temperature 
through the quantum hardening effect on the potential energy. 
The melting temperature $T_m$ is also estimated for full 
understanding the behavior of the solids and for the evaluating 
the temperature range of the new phase. 

Through additional simulation studies, we have also examined to 
see how our conclusion about the existence of the new phase 
is robust against the choice of $V_{\rm pair}$. 
We have found that both $T_{tr}$ and $T_m$ are rather sensitive 
to the change of $V_{\rm pair}$. 
An example of $T_{tr}$ is given 
in Fig.~\ref{fig:referphased}(b).
However, there is always a melting transition {\it after}
 the I/I' transition with increasing temperature, 
suggesting the robustness of the existence of the $Cmca$ phase. 

\acknowledgments

This work is supported by Chinese National Foundation for 
Doctoral Education, Scientific Research Foundation for 
Returned Oversee Chinese Scholars, 
Chinese National Natural Science Foundation, 
and Japanese Monbusho Scholarship for Exchange 
Visiting Researchers. 


\end{multicols}


\begin{references}
\bibitem{anharmonicity}In this toy model, the zero-point 
oscillation is considered in a harmonic potential due to the 
simplifying fact that an analytic expression for $\psi_{0}(x)$ 
is known. 
Even if it cannot be described well by a harmonic potential, 
our conclusion still holds; 
if anharmonicity plays a role in the zero-point oscillation, 
we can easily include its effect by redefining $\psi_{0}(x)$ 
as the ground-state wavefunction for the system 
with the anharmonicity, leading to an equation in just the same form 
as in Eq.~(\ref{eq:toyfinal}) with $\tilde{V}(x)$ defined by 
the first equation in Eq.~(\ref{eq:effectv}). 
(Note that the second equation does not hold in this 
anharmonic case.)
\end{references}
\end{document}